\documentclass[12pt]{article}
\title{A model and characterization of a class of
    \\
    symmetric semibounded operators}
\author{M.\,I.\,Belishev\thanks{St. Petersburg Department of Steklov Mathematical Institute of Russian Academy of Sciences, Fontanka 27, St. Petersburg, Russia, 191023; belishev@pdmi.ras.ru},\and
    S.\,A.\,Simonov\thanks{St. Petersburg Department of Steklov Mathematical Institute of Russian Academy of Sciences, Fontanka 27, St. Petersburg, Russia, 191023; Alferov Academic University of the Russian Academy of Sciences, Khlopina 8/3, St. Petersburg, Russia; sergey.a.simonov@gmail.com}}
\date{}

\usepackage[utf8]{inputenc}
\usepackage{easyReview}
\usepackage{ulem}
\usepackage{amssymb,amsmath,amsthm}
\usepackage{graphicx,color}
\usepackage{mathrsfs}
\usepackage{array, amsfonts, mathrsfs}
\usepackage{epstopdf}
\usepackage[colorlinks=true]{hyperref}
\usepackage{float}
\usepackage{enumitem}
\newtheorem{Lemma}{Lemma}
\newtheorem{Theorem}{Theorem}
\newtheorem{Condition}{Condition}

\newtheorem{Proposition}{Proposition}
\newtheorem*{Definition*}{Definition}

\def\matr{\mathbb M^n_{\mathbb C}}

\def\red{\color{red}}

\def\mFTs{\mathscr F^T_s}
\def\mFT{\mathscr F^{T}}

\def\mF{\mathscr F}
\def\mG{\mathscr G}
\def\mH{\mathscr H}
\def\mK{\mathscr K}

\def\mU{\mathscr U}
\def\mUs{{\mathscr U}^s}
\def\mUT{{\mathscr U}^T}

\def\fB{\mathfrak B}

\def\Dom{{\rm Dom\,}}
\def\Ker{{\rm Ker\,}}

\def\supp{{\rm supp\,}}
\def\dx{\frac{d^2}{dx^2}}

\def\dtau{\frac{d^2}{d\tau^2}}

\def\LT{L_0^{*\,T}}

\def\tLT{\tilde L_0^{*\,T}}

\def\dUT{\dot{\mathscr U}^T}
\def\tU{\tilde{\mathscr U}}

\def\tdUab{\tilde{\mU}^{ab}_m}

\def\dFT{\dot{\mathscr F^T}}

\def\harp{\upharpoonright}
\def\ls{\leqslant}
\def\gs{\geqslant}
\def\bul{\noindent$\bullet$\,\,\,}
\def\rmv{_{\rm v}}

\begin{document}
\maketitle

\begin{abstract}
Let $\mG$ be a Hilbert space and $\fB(\mG)$ the algebra of bounded
operators, ${ \mH}=L_2([0,\infty);\mG)$. An
operator-valued function $Q\in L_{\infty,\rm
loc}\left([0,\infty);\fB(\mG)\right)$ determines a multiplication
operator in $\mH$ by $(Qy)(x)=Q(x)y(x)$, $x\geqslant0$. We say
that an operator $L_0$ in a Hilbert space is a Schr\"odinger
type operator, if it is unitarily equivalent to $-\dx+Q(x)$ on a
relevant domain. The paper provides a characterization of a class
of such operators. The characterization is given in terms of
properties of an evolutionary  dynamical system associated with
$L_0$. It provides a way to construct a functional
Schr\"odinger model of $L_0$.
\end{abstract}

\subsubsection*{About the paper}

\bul Let $\mG$ be a (separable) Hilbert space, $\fB(\mG)$
the algebra of bounded operators, ${\mH}:=L_2([0,\infty);\mG)$. A
locally bounded operator-valued function $Q\in L_{\infty,\rm
loc}\left([0,\infty);\fB(\mG)\right)$ determines the operator of
multiplication in $\mH$ by the rule $(Qy)(x)=Q(x)y(x)$,
$x\geqslant0$. We call an operator in a Hilbert space an {\it
Schr\"odinger type operator}, if it is unitarily equivalent to a
Schr\"odinger operator $-\dx+Q(x)$ on a relevant domain in
$\mH$. Our paper provides a characterization of a class of such
operators. The characterization is given in terms of some
properties of a dynamical system associated with these operators.
It provides a way to construct a functional Schr\"odinger model of a Schr\"odinger type
operator.
\smallskip

\bul The system used for constructing the model is a second-order evolutionary dynamical system.
In many applications, it is governed by various versions of the (hyperbolic) wave equation, which motivates the terminology we use: waves, a wave model, wave subspaces, and so on.

The wave model owes its appearance to inverse problems of
mathematical physics. There is an approach to inverse problems,
the so-called {\it Boundary Control method}, based on their deep
relations with control and system theories, functional analysis,
operator theory. Its achievements include reconstruction of the
Riemannian manifold from spectral and dynamical inverse data
\cite{B UMN,B IPI,BSim_Mat_Sbor}. At some point it became clear
that solving the problem by the BC-method is in fact equivalent to
constructing a functional model of the operator that determines
the evolution of a relevant dynamical system. Such a theoretical
background is revealed and analyzed in
\cite{BSim_FAN,BSim_Mat_Sbor,BSim_A&A_2024}.

The novelty and advantage of the wave model may be demonstrated by
the following example. Assume that we are given the characteristic
function (or the Weyl--Titchmarsh function) of a minimal
Schr\"odinger operator $L_0=-\dx+q(x)$ in $L_2([0,\infty))$.
Constructing traditional models (see, e.\,g., \cite{Nagy-Foias, Pavlov, Naboko, Brodskii-Livsic-1958, Derkach-Malamud-1995, Malamud-Malamud-2003, Strauss}, we can realize $L_0$ as the
multiplication operator by $z\in\Omega\subset\mathbb C$ on
holomorphic functions $f(z)$, which take values in a relevant
Hilbert space. At the same time, constructing the {\it wave
model}, we get the operator $-\dx+q(x)$ \cite{BSim_A&A_2017}. This
clarifies usefulness of the wave model and its productivity in
applications. However, of course, in contrast to known general
models, the wave model is relevant for a narrower specific class
of operators. In particular, the semi-boundedness (the positive
definiteness) of $L_0$ is substantial.
\smallskip

\bul A rather short list of references to papers dealing with models is explained by the fact that we did not find any real predecessors of the wave model in the literature.

\subsubsection*{Operators and systems}

\noindent$\bullet$\,\,\,Let $L_0$ be a closed symmetric positive definite operator in a Hilbert space ${\mathscr H}$ with the defect indices $1\ls n_\pm^{L_0}\ls\infty$.
Let $L$ be the extension of $L_0$ by Friedrichs, so that
\begin{equation}\label{Eq L0,L,L*}
    L_0\subset L\subset L_0^*
\end{equation}
holds. Let $P$ be the projection in $\mH$ onto $\mK:={\rm Ker\,}L_0^*$. From the assumptions, $L^{-1}=(L^{-1})^*\in\fB(\mH)$.

The operators $\Gamma_1:=L^{-1}{L_0^*}-\mathbb I$, $\Gamma_2:=P{L_0^*}$, ${\rm Dom\,}\Gamma_{1,2}={\rm
Dom\,}L_0^*$, ${\rm Ran\,}\Gamma_{1,2}=\mathscr K$, are called the {\it boundary operators}. The Green formula
\begin{equation*}
    ({L_0^*} u,v)-(u,{L_0^*}v)=(\Gamma_1u,\Gamma_2v)-(\Gamma_2u,\Gamma_1v), \qquad u,v\in{\rm Dom\,}L_0^*,
\end{equation*}
holds. For operators in (\ref{Eq L0,L,L*}) one has
\begin{equation*}
    L_0=L^*_0\harp\left[\Ker\Gamma_1\cap\Ker\Gamma_2\right],\quad L=L^*_0\harp\Ker\Gamma_1.
\end{equation*}
The well-known M.\,I.\,Vishik decompositions are
\begin{equation*}\label{Eq Vishik decomp}
    \Dom L^*_0=\Dom L_0\dotplus L^{-1}\mK\dotplus\mK=\Dom L\dotplus\mK,\quad\Dom L=\Dom L_0\dotplus L^{-1}\mK
\end{equation*}
(see, e.\,g., \cite{Vishik,BD_DSBC,MMM}).
The triple $(\mK; \Gamma_1, \Gamma_2)$ is an {\it ordinary boundary triple} for the operator $L_0^*$ \cite{MMM}.
\smallskip

\noindent$\bullet$\,\,\,Fix $T>0$. The operator $L_0$ determines the dynamical system $\alpha^T$ of the form
\begin{align}
    \label{Eq 1}& u''(t)+L_0^*u(t) = 0  && {\rm in}\,\,\,{{\mathscr H}}, \,\,\,t\in(0,T),\\
    \label{Eq 2}& u(0)=u'(0)=0 && {\rm in}\,\,\,{{\mathscr H}},\\
    \label{Eq 3}&\Gamma_1 u(t) = f(t) && {\rm in}\,\,\,{{\mathscr K}},\,\,\,t\in[0,T],
\end{align}
where a ${\mathscr K}$-valued function of time $f=f(t)$ is a {\it boundary control}, $u=u^f(t)$ is the solution (a {\it
wave}). System theory attributes of $\alpha^T$ are as follows.

\noindent{\bf 1.}\,\,\,The {\it outer space} of controls is $\mFT:=L_2([0,T];\mK)$. The class of {\it smooth controls}
$\dot\mFT:=\{f\in C^\infty([0,T];\mK)\,|\,\,{\rm supp\,}f\subset(0,T]\}$ is dense in $\mFT$ and satisfies
\begin{equation}\label{Eq d^pF=F}
    \frac{d^{\,p}}{dt^{\,p}}\,\dFT=\dFT,\qquad p=1,2,\dots\,.
\end{equation}
For $f\in\dot\mF^T$ the classical solution $u^f$ is unique and the relation
\begin{equation}\label{Eq u in Dom L*0}
    u^f(t)\in{\rm Dom\,}L^*_0,\qquad t>0,
\end{equation}
holds. Representations
%\begin{multline}\label{Eq repres u^f 0}
%u^f(t)=-f(t)+L^{-{\frac12}}\int_0^t\sin[(t-s)L^{\frac12}]\,f''(s)\,ds\\
%    =-f(t)+\int_0^t\cos[(t-s)L^{\frac12}]\,f'(s)\,ds\\
%=-f(t)+L^{-1}\int_0^t\left(\mathbb
%I-\cos[(t-s)L^{\frac12}]\right)\,f'''(s)\,ds
%\end{multline}
\begin{align}
\notag & u^f(t)=-f(t)+L^{-{\frac12}}\int_0^t\sin[(t-s)L^{\frac12}]\,f''(s)\,ds\\
\notag &    =-f(t)+\int_0^t\cos[(t-s)L^{\frac12}]\,f'(s)\,ds\\
\label{Eq repres u^f 0} & =-f(t)+L^{-1}\int_0^t\left(\mathbb
I-\cos[(t-s)L^{\frac12}]\right)\,f'''(s)\,ds
\end{align}
for $f\in\dot\mF^T$ take place \cite{BD_DSBC}. Here equalities are
derived using integration by parts.

Since the operator $L_0^*$ that governs the evolution of
$\alpha^T$ does not depend on time, the equalities
\begin{equation}\label{Eq steady state}
\mathcal{}u^{-f''}(t)=-(u^f)''\overset{(\ref{Eq 1})}=L_0^*\,u^f(t),\qquad t>0,
\end{equation}
hold. The space $\mFT$ contains the extending family (a {\it
nest}) of subspaces of delayed controls
%$$
%\mFTs:=\{f\in\mF\,|\,\,f\upharpoonright\{0\ls t\ls T-s\}\equiv0\},\qquad s\in[0,T];
%$$
$$
\mFTs:=\{f\in\mF\,|\,\,{\rm
supp\,}f\subset[T-s,T]\}\},\qquad s\in[0,T];
$$
here $s$ is the time of action and $T-s$ is the delay, so that
$\mF^T_0=\{0\}$ and $\mF^T_T=\mF^T$ holds. We put
$\dot\mFTs:=\mFTs\cap{\dot\mFT}$.
\smallskip

\noindent{\bf 2.}\,\,\,The {\it inner space} of states is $\mH$.
It contains the nest of {\it reachable sets}
$$
\dot\mUs:=\{u^f(s)\,|\,\,f\in\dot\mFT\},\qquad s\in [0,T];
$$
we call the elements of $\dot\mU^s$ the {\it smooth waves}. Note
that the definition of $\dot\mUs$ does not depend on $T$. The
invariance (\ref{Eq d^pF=F}) and relations \eqref{Eq u in Dom
L*0}, \eqref{Eq steady state} lead to the equality
\begin{equation*}\label{Eq invariance}
    L_0^*\,\dot\mUs\,=\,\dot\mUs,\qquad s\in [0,T]\,.
\end{equation*}
We denote $\mU^s:=\overline{\dot \mU^s}$ and call it the {\it wave subspace}.
\smallskip

\noindent{\bf 3.}\,\,\, The {\it control operator}
$W^T:\mFT\to\mH$,\,\,$W^Tf:=u^f(T)$, is defined on $\dot\mFT$. It
can be unbounded, but is always closable \cite{B DSBC_3}. The
second of the representations \eqref{Eq repres u^f 0} shows that
$W^T$ can be extended from $\dot\mF^T$ to the Sobolev space
$\mF^T_1:=\{f\in W^1_1([0,T];\mK)\,|\,\,f(0)=0\}$ so that the
extension is a bounded operator from $\mF^T_1$ to $\mH$. Closure
of $W^T$ is also denoted by $W^T$. We have $\dot\mU^T=W^T\dot\mFT$
and $\mU^T=\overline{W^T\dot\mFT}$.
\smallskip

\noindent{\bf 4.}\,\,\, By the von Neumann theorem, the operator
$C^T:=(W^T)^*W^T$ is densely defined and positive (but not
necessarily positive definite) in $\mFT$, whereas its closure
(also denoted by $C^T$) satisfies $C^T=(C^T)^*$ \cite{BirSol}. It
connects the metrics of the outer and the inner spaces by the
equalities
\begin{equation*}\label{Eq C^T connects}
    (C^Tf,g)_\mFT=(W^Tf,W^Tg)_\mH=(u^f(T),u^g(T))_\mH,\qquad f,g \in {\rm Dom\,}C^T,
\end{equation*}
and is called {\red a} {\it connecting operator}.
\smallskip

\subsubsection*{Operator parts}

\bul The operator $\dot L_{0}^{*\,T}:={L_0^*\harp\dot\mU^T}$ is
densely defined in $\mU^T$ and can also be defined by its graph
\begin{equation*}\label{Eq graph}
    {\rm graph\,}\dot L_{0}^{*\,T}\,=\,{\{(W^Tf,-W^Tf'')\,|\,\,f\in\dot\mF^T\}}.
\end{equation*}
Introduce the total reachable set $\dot\mU:={\rm
span}\,\{\dUT\,|\,\,T>0\}$ and note its invariance
$L^*_0\dot\mU=\dot\mU$. The subspace
\begin{equation*}
\mU:=\overline{\dot\mU}\subset\mH
\end{equation*}
is called the {\it total wave subspace}.

Let $\mG$ and $\mG'\subset\mG$ be a Hilbert space and its (closed)
subspace, let $A$ be an operator in $\mG$. The subspace $\mG'$ is
called an {\it invariant subspace} of $A$, if
$$
    \overline{\mG'\cap{\rm Dom\,}A}=\mG',\qquad A\,[\mG'\cap{\rm Dom\,}A]\subset\mG'
$$
holds \cite{BSim_ZNS_2019}. The operator
$A_{\mG'}:=A\harp[\mG'\cap{\rm Dom\,}A]: \mG'\to\mG'$ is called
the {\it part} of $A$ in $\mG'$. The part is necessarily a closed
operator. 

The subspace $\mG'$ {\it splits} the operator $A$, if the subspaces $\mG'$ and $\mG\ominus\mG'$ are invariant for it. If additionally $P_{\mG'}\Dom A=\Dom A\cap\mG'$, then the subspace $\mG'$ {\it reduces} the operator $A$. It is known that every
symmetric non-self-adjoint operator has the smallest reducing
subspace such that its part there is non-self-adjoint (the {\it
completely non-self-adjoint}, or {\it simple}, part); the part of
the operator in the orthogonal complement (which may be trivial)
to that subspace is self-adjoint.

%If a subspace {\it reduces} $A$, it, in particular, is invariant for $A$, and the part of $A$ in this reducing subspace is the same as defined above.

In \cite{BD_DSBC}, the following is shown.

\begin{Proposition}\label{Prop 0}
The subspace $\mU$ reduces the symmetric operator $L_0$,
and the part of ${L_0}_{\mU}$ is its completely non-self-adjoint
part.
\end{Proposition}

\noindent Hence, if $L_0$ is completely non-self-adjoint, then $\mU=\mH$.

\begin{Definition*}\label{Def wave parts}
The operators $L^{*\,T}_0:=\overline{\dot L_{0}^{*\,T}}$ and
$L^{*\,\infty}_0:=\overline{L^*_0\harp\dot\mU}$ are called the
wave part of $L^*_0$ for the time $T$ and the wave part of
$L^*_0$, respectively.
\end{Definition*}

The equality $L^{*\,\infty}_0={L^*_0}_{\mU}$ is not guaranteed.
Note that the case $\mU^T=\mU$ and
$L^{*\,T}_0=L^{*\,\infty}_0={L^*_0}_{\mU}$ for all $T>0$ is
possible, but is not interesting \cite{B DSBC_3}.
\smallskip

\bul Here we formulate the first of the conditions on the operator
$L_0$, which provide a characterization of the class of
Schr\"odinger type operators that we consider. We begin with an inspiring example
of an invariant subspace.

Let $\mG=L_2([0,\infty);\mathbb C^n)$; $C_{\rm
c}^{\infty}((0,\infty);\mathbb C^n)\subset\mG$ is the class of smooth
vector-functions compactly supported in $(0,\infty)$. Assume that
$q=q(x)$ is a locally bounded Hermitian matrix-valued function
such that the operator
$$
S_0:=\overline{\left(-\dx +q\right)\upharpoonright C_{\rm
c}^{\infty}((0,\infty);\mathbb C^n)}
$$
is positive definite. Then from the results of
\cite{Povzner-Wienholz} it follows that the adjoint of this
operator acts by $S_0^*=-\dx +q(x)$ on the domain
$$
\Dom S_0^*=\{y\in L_2([0,\infty);\mathbb C^n)\cap H^2_{\rm loc}([0,\infty);\mathbb C^n)\,|\,\,-y''+qy\in L_2([0,\infty);\mathbb C^n)\},
$$
and the operator $S_0$ acts on the domain
$$
\Dom S_0 = \{y\in \Dom S_0^*\,|\,y(0)=y'(0)=0\}
$$
and is symmetric with defect indices $n_{\pm}^{S_0}=n$. The
subspaces $\mG^{ab}:=\{y\in\mG\,|\,\,\supp
y\subset[a,b]\subset[0,\infty)\}$,\,\,$0\ls a<b<\infty$, are
invariant for both $S_0$ and $S_0^*$. Moreover, in the case
$0<a<b<\infty$ one has $S_{0\,\mG^{ab}}=S_{0\,\mG^{ab}}^*$, so
that the part $S_{0\,\mG^{ab}}^*$ is a symmetric operator. For
$0\ls a<b<\infty$ we put
$$
\mU^{ab}:=\mU^{b}\ominus\mU^a.
$$
It turns out (see the next section) that the relations
$\mU^T=L_2([0,T];\mathbb C^n)$, $\mU^{ab}=\mG^{ab}$, and
$$
\Dom S_0^{*\,T}=\{y\in H^2([0,T];\mathbb C^n)\,|\,y(T)=y'(T)=0\},
$$
are valid, whereas $\mU^{ab}$ is an invariant subspace for
$S_0^{*\,T}$ and $\Dom S_{0\,\mU^{ab}}^{*\,T}=\Dom
S_{0\,\mG^{ab}}^*$ hold.
\smallskip

\bul The following assumptions give a relevant abstract
version of the Schr\"odinger operator properties mentioned
above.

\begin{Condition}\label{C L0 split}
For every finite $T>0$ and $0\ls a<b\ls T$, the subspace
$\mU^{ab}$ is an invariant subspace of the operator $L_0^{*\,T}$.
If $0<a<b\leqslant T$, then the part $L_{0\,\mU^{ab}}^{*\,T}$ is a
symmetric operator.
\end{Condition}

Regarding the first part of Condition \ref{C L0 split}, it is
worth to note the following fact. If the subspace $\mU^T$ is invariant for $L_0^*$, then the wave part $L_0^{*\,T}$ and
the part $L_{0\,\mU^T}^*$ (the {\it space part}, cf. \cite{B
DSBC_3}) are related as $L_0^{*\,T}\subset
L_{0\,\mU^T}^*$, and their coincidence may not hold. However, the
following is shown in \cite{B DSBC_3}. By an {\it isomorphism} we
mean a bounded and boundedly invertible operator.

\begin{Proposition}\label{P0}
If $W^T:\mF^T\to\mU^T$ is an isomorphism, the subspace $\mU^T$ is invariant for the operator $L_0^*$ and the relation
$\mU^T\cap\mK=\{0\}$ holds, then the equality
$L_0^{*\,T}=L_{0\,\mU^T}^*$ is valid.
\end{Proposition}

It is possible that the second part of Condition \ref{C L0 split}
can be derived from general properties of the system $\alpha^T$: a
close statement is established in \cite{B DSBC_3}, Lemma 9.

\subsubsection*{The diagonal}

\bul Let $\mF$ and $\mH$ be two Hilbert spaces and $\mathfrak
f=\{\mF_s\}_{0\ls s\ls T}$ be a nest of subspaces in $\mF$ obeying
$\{0\}=\mF_0\subset \mF_s\subset\mF_{s'}\subset
\mF_T=\mF,\,\,\,s<s'$. Let $X_s$ be the projection in $\mF$ onto
$\mF_s$. For a bounded operator $A:\mF\to\mH$ by $P_s$ we denote
the projection in $\mH$ onto $\overline{A\mF_s}$. Choose a
partition $\Xi=\{s_k\}_{k=0}^N:\quad 0=s_0<s_1<\dots<s_N=T$ of
$[0,T]$ of the range
$r^{\Xi}:=\max\limits_{k=1,\dots,N}(s_k-s_{k-1})$. Denote $\Delta
X_k:=X_{s_k}-X_{s_{k-1}}$, $\Delta P_k:=P_{s_k}-P_{s_{k-1}}$ and
put \begin{equation}\label{Eq sums}
    D^{\Xi}_A\,:=\,\sum\limits_{k=1}^N\Delta P_k\,A\,\Delta X_k.
\end{equation}
The operator $D_{A}: {\mathscr F}\to{\mathscr H}$,
$D_A=\text{w\,-}\!\lim\limits_{r^\Xi\to
0}\,D^{\Xi}_A=:\int_{[0,T]}dP_s\,A\,dX_s$ is called the {\it
diagonal} of $A$ with respect to the nest $\mathfrak f$. Not every
isomorphism possesses a diagonal (A.\,B.\,Pushnitskii,
\cite{BPush}).

If it exists, the diagonal intertwines the projections:
$P_sD_{A}=D_{A}X_s$ holds for all $s$. The representation
$D^*_A=\int_{[0,T]}dX_s\,A^*\,dP_s$ is valid. Construction of the
diagonal generalizes the classical triangular truncation integral
by M.\,S.\,Brodskii and M.\,G.\,Krein \cite{Brod,GK,B Obzor IP
97,BSim_A&A_2024}.
\smallskip

\bul For the system $\alpha^T$ take the nest $\mathfrak
f^T=\{\mF^T_s\}_{0\ls s\ls T}$; let $X^T_s$ and $P_s$ be the
projections in $\mFT$ onto $\mFTs$ and in $\mU^T$ onto ${\mUs}$,
respectively. The following assumption on the control operator is
in fact an assumption imposed implicitly on the operator $L_0$,
which determines the system (\ref{Eq 1})--(\ref{Eq 3}).
\begin{Condition}\label{C2}
For every $T>0$ the operator $W^T:\mFT\to\mU^T$ is
bounded and injective. It possesses the diagonal
$D_{W^T}=\int_{[0,T]}dP_s\,W^T\,dX^T_s$ obeying $\Ker
D_{W^T}=\{0\}$ and $\overline{{\rm Ran\,} D_{W^T}}=\mU^T$.
\end{Condition}
By terminology of \cite{BSim_A&A_2024}, $W^T$ is a {\rm
strongly regular} operator.

\noindent This condition is inspired by applications to inverse
problems \cite{B Obzor IP 97,B IPI}. Under such assumptions, the
following is proved in \cite{BSim_A&A_2024}.

The connecting operator $C^T=(W^T)^*W^T$ admits a triangular
factorization in $\mFT$ of the form $C^T=(V^T)^*V^T$ with
$$
V^T:=\Phi_{D^*_{W^T}}W^T
$$
obeying $V^T\mFTs=\mFTs$, $s\in [0,T]$, where
$\Phi_{D^*_{W^T}}:\mU^T\to\mF^T$ is the unitary factor in the
polar decomposition $D^*_{W^T}=\Phi_{D^*_{W^T}}|D^*_{W^T}|$. If $W^T$ is an
isomorphism, then the operator $V^T$ is also an isomorphism. The
representation
\begin{equation}\label{Eq sqrt C^T}
V^T=\Phi_{D^*_{\sqrt{C^T}}}\sqrt{C^T}
\end{equation}
holds, where $\sqrt{C^T}$ is the positive square root of $C^T$.
\smallskip

\bul The diagonal realizes the spectral theorem for the
{\it eikonal operator} $E^T:=\int_{[0,T]}s\,dP_s$, which is
self-adjoint and positive in $\mU^T$: the relation
\begin{equation}\label{Eq Eikonal diag}
    \hat E^T:=\Phi_{D_{W^T}^*}E^T\,(\Phi_{D^*_{W^T}})^*\,=\,\int_{[0,T]}s\,dX^T_s\,=\,T\,\mathbb
    I-\hat t
\end{equation}
holds, where $\mathbb I$ is the identity operator in $\mF^T$ and
$\hat t$ is the multiplication by the variable $t$
(the time): $(\hat tf)(t)=tf(t)$, $0\ls t\ls T$,
\cite{BSim_A&A_2024}.

\subsubsection*{Models}

\bul Introduce the {\it model space}
$\tilde\mU:=L_2([0,\infty);\mK)$ of $\mK$-valued functions
$y=y(\tau)$, $\tau>0$, and its subspaces
$\tilde{\mU^T}:=\{y\in\tU\,|\,\,{\rm
supp\,}y\subset[0,T]\}=L_2([0,T];\mK)$. We use the
auxiliary operators  $Y^T:\mFT\to\mFT,
\,(Y^Tf)(t):=f(T-t)$,\,\,$0\ls t\ls T$ and $\tilde
Y^T:\mFT\to\tilde\mU^T$,\, $(\tilde
Y^Tf)(\tau):=f(T-\tau)$,\,\,$0\ls \tau\ls T$. Define the {\it
model control operator} $\tilde W^T:\mFT\to\tilde\mUT$,
$$
\tilde W^T:=\tilde Y^TV^TY^T=\Phi^T W^TY^T
$$
with the unitary (under Condition \ref{C2}) map $\Phi^T:=\tilde
Y^T\Phi_{D^*_{W^T}}$ from $\mU^T$ to $\tilde\mU^T$. According to the results of \cite{BSim_A&A_2024},
the families $\{\tilde W^T\}_{T>0}$ and $\{\Phi^T\}_{T>0}$ possess
the property
\begin{equation*}\label{continuation property}
    \tilde W^T=\tilde W^{T'}\upharpoonright\mFT,\quad \Phi^T=\Phi^{T'}\upharpoonright\mUT,\quad T<T'.
\end{equation*}
Moreover, there exists a unitary operator $\Phi:\mU\to\tilde\mU$
(the so-called {\it global orthogonalizer}) such that
$$
\Phi^T=\Phi\upharpoonright\mU^T,\quad T>0
$$
holds.

The wave part of the operator $L_0^*$ and all its parts are
transferred to the model space $\tilde\mU$:  operators
$$
\tilde L_{0_{\mU}}^*:=\Phi L_{0_{\mU}}^*\Phi^*, \quad\tilde L_0^{*\,T}:=\Phi L_0^{*\,T}\Phi^*.
$$
are regarded as models of $L_{0_{\mU}}^*$ and $L_0^{*\,T}$,
respectively. The following assumption is imposed on smoothness
of functions from $\Dom\tilde L_0^{*\,T}=\Phi\Dom L_0^{*\,T}$.
\begin{Condition}\label{C3}
For every $T>0$ the inclusion ${\rm Dom\,}\tilde
{L}_{0}^{*\,T}\subset H^2([0,T];\mK)$ holds.
\end{Condition}

\noindent By the latter, the operator
$$
Q^T:=\tilde {L}_{0}^{*\,T}+\dtau
$$
in $\tilde\mU^T$ is defined on ${\rm Dom\,}\tilde {L}_{0}^{*\,T}$.
\smallskip

\begin{Condition}\label{C4}
For every $T>0$ the operator $Q^T$ is bounded.
\end{Condition}

\bul The conditions accepted above are motivated by the
following result.

\begin{Lemma}\label{lemma 1}
Under Conditions \ref{C L0 split}--\ref{C4} there exists an
operator-valued function $q\in L_{\infty,{\rm
loc}}([0,\infty);\fB(\mK))$ such that $q(\tau)=q^*(\tau)$ holds for every $\tau\gs 0$, and for every $T>0$, $y\in
L_2([0,T];\mK)$ one has $(Q^Ty)(\tau)=q(\tau)y(\tau)$,
$\tau\in[0,T]$. In other words, $Q^T$ is a self-adjoint {\rm
decomposable} operator in $\tilde\mUT=L_2([0,T];\mK)$.
\end{Lemma}

\begin{proof}
Let $0<a<b<T$ and
$\tU^{ab}:=\Phi\mU^{ab}=\Phi(\mU^b\ominus\mU^a)=(\Phi\mU^b)\ominus(\Phi\mU^a)=\tilde\mU^b\ominus\tilde\mU^a=L_2([a,b];\mK)$.
Consider the linear set \begin{equation*}
\tdUab:=\Dom\tLT\cap\tU^{ab}\subset H^2([0,T];\mK)\cap
L_2([a,b];\mK).
\end{equation*}
For every $y\in\tdUab$ one has $y(a)=y'(a)=y(b)=y'(b)=0$, so
$\tdUab\subset\mathring H^2([a,b];\mK)$. Owing to Condition \ref{C
L0 split} and the unitarity of $\Phi$, $\tdUab=\Phi(\Dom
L_0^{*\,T}\cap\mU^{ab})$ is dense in $\tilde\mU^{ab}$, hence the
operator $\frac{d^2}{d\tau^2}\upharpoonright\tdUab$ is symmetric
in $\tU^{ab}$. By the same Condition \ref{C L0 split}, the
restriction $\tLT\upharpoonright\tdUab=\tilde
L^{*\,T}_{0_{\mU^{ab}}}=\Phi(L^{*\,T}_{0_{\mU^{ab}}})\Phi^*$ is
also symmetric in $\tU^{ab}$, and hence such is
$Q^T\upharpoonright\tdUab$. The linear span of $\tdUab$ over all
$a,b$ such that $0<a<b<T$ is dense in $\tU^T$, therefore
$\overline{Q^T}$ is a bounded self-adjoint operator.

Let us show that the subspaces $\tU^{ab}$, $0\ls a<b\ls T$, reduce
the operator $\overline{Q^T}$. From the invariance of $\mU^{ab}$
for $\LT$ and the unitarity of $\Phi$ it follows that for every
$y\in\tdUab$ one has $\tLT y\in\tU^{ab}$; besides that clearly
$y''\in\tU^{ab}$. Therefore $Q^T\tdUab\subset\tU^{ab}$, and hence
$\overline{Q^T}\tU^{ab}\subset\tU^{ab}$. Since $\overline{Q^T}$ is
self-adjoint, this means that the subspace $\tU^{ab}$ is reducing
for $\overline{Q^T}$. We have shown that
$\overline{Q^T}P_{\tU^{ab}}=P_{\tU^{ab}}\overline{Q^T}$ for
$0<a<b<T$. This equality can be extended to the case $0\ls a<b\ls
T$ by taking a limit in the sense of strong operator convergence.

Consider the space $\tU^T=L_2([0,T];\mK)$ as a direct integral of
Hilbert spaces $\mK$, i.\,e., $\tU^T=\oplus\int_{[0,T]}\mK d\tau$.
The projection-valued measure $d\tilde P_{\tau}$, where $\tilde
P_{\tau}:=P_{\tU^{\tau}}$, is the spectral measure of the operator
$[\tau]$ of multiplication by the independent variable in this
space. By \cite[Theorem 7.2.3]{BirSol}, commutation
\begin{equation}\label{kommutatsiya}
\overline{Q^T}\tilde P(\delta)=\tilde P(\delta)\overline{Q^T}
\end{equation}
for every Borel set $\delta$ implies {\it decomposability} of
$\overline{Q^T}$: there exists an operator-valued function $q^T\in
L_{\infty}([0,T];\fB(\mK))$ such that
$(\overline{Q^T}y)(\tau)=q^T(\tau)y(\tau)$ for a.\,e.
$\tau\in[0,T]$ and
$\|q^T\|_{L_{\infty}([0,T];\fB(\mK))}=\|\overline{Q^T}\|_{\fB(\tU^T)}$.
One can show that since the measure in the direct integral is the
Lebesgue measure, the condition \eqref{kommutatsiya} can be
checked only for intervals $\delta=(a,b)$, $0\leqslant
a<b\leqslant T$, and it holds for intervals in our case. The
property of the family of operators $\overline{Q^T}$,
$$
\overline{Q^T}=\overline{Q^{T'}}\upharpoonright\tU^T,\quad T'>T,
$$
implies that
$$
q^T=q^{T'}\upharpoonright[0,T],\quad T'>T,
$$
which means that there exists a function $q\in L_{\infty,\rm
loc}([0,\infty);\fB(\mK))$ such that $q^T=q\upharpoonright[0,T]$
for every $T>0$.
\end{proof}

As a result, we conclude that the model of the wave part
${L}_{0}^{*\,T}$ has the form $\tilde
{L}_{0}^{*\,T}=-\dtau+q(\tau)$, i.\,e., is a Schr\"odinger
operator on some domain in $L_2([0,T];\mK)$. Moreover, the
construction of the model provides an efficient way to
realize $L^{*\,T}_0$ in such a form. To this end, it suffices to
have the connecting operator $C^T$, to provide its
factorization $C^T=(V^T)^*V^T$, determine $\tilde W^T$ and
then to find the model $\tilde{L}_{0}^{*\,T}$ via its graph
\begin{equation*}\label{Eq graph}
    {\rm graph\,}\tilde{L}_{0}^{*\,T}\,=\,\overline{{\{(\tilde W^Tf,-\tilde W^Tf'')\,|\,\,f\in\dot\mF^T\}}}.
\end{equation*}
A remarkable fact is that in actual applications the inverse
data determine the connecting operator. The latter enables one to
recover the `potential' $q$ from the data. We may call the
operators $\tilde{L}_{0}^{*\,T}$, $T>0$, the {\it local wave
models} of the operator $L_0^*$.
\smallskip

\bul For $0<T<T'$ we evidently have $\tilde
L_0^{*\,T}\subset\tilde L_0^{*\,T'}\subset\tilde L_0^{*\,\infty}$.
Sending $T$ to the infinity, we obtain an extending family of
operators and determine the operator $ \tilde
L_0^{*\,\infty}\upharpoonright{{\rm span\,}_{T>0}\Dom\tilde
L_0^{*\,T}} $ which, after taking the closure, becomes $\tilde
L_0^{*\,\infty}$, the model of the wave part of $L_0^*$. By
construction, this model is a Schr\"odinger operator of the
form $-\dtau+q(\tau)$ acting on a certain domain. With this
differential expression we associate two `standard' Schr\"odinger
operators, defined by their domains: the {\it minimal} $S_{\rm
min}^q$ acting on
$$
\Dom S_{\rm min}^q:=\Dom\left
(\overline{\left[-\frac{d^2}{d\tau^2}+q\right]\upharpoonright
C_{\rm c}^{\infty}((0,\infty);\mK)}\right ),
$$
and the {\it maximal} $S_{\rm max}^q$ acting on
$$
\Dom S_{\rm max}^q:=\{y\in L_2([0,\infty);\mK)\cap H^2_{\rm loc}([0,\infty);\mK)\,|\,-y''+qy\in L_2([0,\infty);\mK)\}.
$$
\smallskip
The model of the wave part $\tilde L_0^{*\,\infty}$ acts on the
domain which is contained in $\Dom S_{\rm max}^q$, but may be
smaller. We arrive at the following result.

\begin{Lemma}\label{lemma 2}
Let a closed symmetric positive definite operator $L_0$ be such
that Conditions \ref{C L0 split}--\ref{C4} hold. Then the wave
part $L^{*\,\infty}_0$ of its adjoint is unitarily equivalent to a Schr\"odinger operator.
\end{Lemma}

Assume in addition that $L_0$ is a completely non-self-adjoint
operator. Then by Proposition \ref{Prop 0} we have $\mU=\mH$, so
that $L^{*\,\infty}_0$ is a densely defined closed Schr\"odinger type
operator. It is not automatically true that its adjoint
$(L_0^{*\,\infty})^*$ is also a Schr\"odinger type operator, unless we
impose one more condition.

\begin{Condition}\label{C5}
The relation $L_0^{*\,\infty}=L^*_0$ holds.
\end{Condition}

\noindent This implies complete non-self-adjointness of $L_0$,
since it means that $\mH\ominus\mU=\{0\}$. Moreover, then
$L_0=(L_0^{*\,\infty})^*\subset L_0^{*\,\infty}$, and hence $L_0$ is also a
Schr\"odinger type operator, so we conclude the following.

\begin{Theorem}\label{T1}
If a closed symmetric positive definite operator $L_0$
satisfies Conditions \ref{C L0 split}--\ref{C5} then its adjoint
$L_0^*$ is unitarily equivalent to a Schr\"odinger operator $-\frac{d^2}{d\tau^2}+q(\tau)$ in $L^2([0,\infty);\mK)$, which is an extension of $S^q_{\rm min}$ and a restriction of $S^q_{\max}$, with an
Hermitian operator-valued potential $q$ from the class
$L_{\infty,\rm loc}([0,\infty);\fB(\mK))$.
\end{Theorem}

\bul The situation becomes significantly simpler, if the defect
indices of the operator $L_0^*$ are finite. In this case the
operator-valued potential becomes equivalent to a matrix-valued
one, and for matrix Schr\"odinger operators an analog of the
Povzner--Wienholtz theorem holds \cite{Povzner-Wienholz}, which
states that positive definiteness of the minimal operator implies
that its defect indices $n_{\pm}^{S_{\rm min}^q}$, which
generically could range from $0$ to $2n$, are in fact
equal to $n$. This means that the defect is related to the
boundary condition at $\tau=0$ and that the maximal and the
minimal operators share the same (absent) boundary condition at
infinity. This leads to the following result. Below $\mathbb
M^n_{\mathbb C}$ denotes square matrices of size $n$ with complex
entries.

\begin{Theorem}\label{T2}
A closed symmetric positive definite operator $L_0$ with finite
defect indices satisfying Conditions \ref{C L0 split}--\ref{C5} is
unitarily equivalent to a minimal Schr\"odinger operator $S^q_{\rm
min}=-\frac{d^2}{d\tau^2}+q(\tau)$ with an Hermitian matrix-valued
potential $q\in L_{\infty,\rm loc}([0,\infty);\mathbb M^n_{\mathbb
C})$.
\end{Theorem}

\begin{proof}
The situation of Theorem \ref{T1} can be immediately reduced from
the $\mK$-valued $L_2$ space to the $\mathbb C^n$-valued one by
picking an orthonormal base $\hat k_1,...,\hat k_n$ in $\mK$ and
taking the unitary transform
$$
L_2([0,\infty);\mK)\ni y(\cdot)\mapsto\hat y(\cdot)=((y(\cdot),\hat k_i)_{\mH})_{i=1}^n\in L_2([0,\infty);\mathbb C^n).
$$
The resulting operator $\hat L_0^*=\hat L_0^{*\,\infty}$ acts as
$-\frac{d^2}{d\tau^2}+\hat q(\tau)$ with a matrix-valued locally
bounded potential $\hat q$ on some domain contained in $\Dom
S_{\rm max}^{\hat q}$. It is known that $S_{\rm max}^{\hat
q}=(S_{\rm min}^{\hat q})^*$, thus one has
$$
S_{\rm min}^{\hat q}\subset\hat L_0\subset \hat L_0^*\subset S_{\rm max}^{\hat q}.
$$
The defect indices of the operators $S_{\rm min}^{\hat q}$ and
$\hat L_0$ coincide, which means that these operators are the
same, and one can take $\hat q$ as $q$ from the statement of the
theorem.
\end{proof}

\bul In the light of the spectral theorem, the unitary operator
$\Phi:\mU\to\tilde\mU$ that provides the wave models to
$L^*_0$ and $L_0$, is a Fourier transform, which diagonalizes the
eikonal operator $E:=\int_{[0,\infty)}t\,dP_t$ by transferring it
to the operator of multiplication by independent variable: $\tilde
E:=\Phi E\Phi^*=\hat\tau$ in $\tilde\mU$, see \eqref{Eq Eikonal
diag}. Such a transform is not unique, but constructing the model
based on factorization (\ref{Eq sqrt C^T}), we select a {\it
canonical} one. From the fact that $\tilde E=\hat\tau$ we conclude
that under Conditions \ref{C L0 split}--\ref{C4} the eikonal $E$
has the spectrum $\sigma(E)=\sigma_{\rm ac}(E)=[0,\infty)$ of
constant multiplicity ${\rm dim\,}\mK$.

\subsubsection*{Characterization}

\bul In what follows we deal with an operator $L_0$ which
satisfies the assumptions of Theorem \ref{T2}. It turns out that
in such a case a characterization takes place.

\begin{Theorem}\label{T3}
Let $L_0$ be a closed symmetric positive definite operator with
finite defect indices. Then $L_0$ is unitarily equivalent to a
minimal Schr\"odinger operator, if and only if it satisfies
Conditions \ref{C L0 split}--\ref{C5}.
\end{Theorem}

\noindent Sufficiency of these conditions is already shown by
Theorem \ref{T2}. To prove necessity, it remains to show that a
minimal matrix Schr\"odinger operator does satisfy Conditions
\ref{C L0 split}--\ref{C5}. Indeed, then for an operator which is
unitarily equivalent to such an operator, these conditions are
fulfilled automatically in view of their invariant character.
\smallskip

\bul In the space $\mH=L_2([0,\infty);\mathbb C^n)$ consider the
minimal Schr\"odinger operator
\begin{equation*}\label{Eq Oper SL}
S_0:=S_{\rm min}^q\,=\,\overline{\left[-\dx+q\right]\harp
C^\infty_{\rm c}\left((0,\infty); {\mathbb C^n}\right)}\,,
\end{equation*}
where $q=q(x)$ is a locally bounded Hermitian $\mathbb
M^n_{\mathbb C}$\,-\,valued function.

\begin{Lemma}\label{L2 final}
If the operator $S_0$ is positive definite, then it satisfies
Conditions \ref{C L0 split}--\ref{C5}.
\end{Lemma}

\begin{proof}
\noindent{\bf 1.}\,\,\,The following are well-known facts about $S_0$.
\smallskip

\noindent$\ast$\,\,\,Assuming that $S_0$ is positive definite, we
denote by $S$ its Friedrichs extension. The following relations
hold by virtue of the Povzner--Wienholz theorem
\cite{Povzner-Wienholz}:
\begin{align*}
& \Dom S_0^* =\{y\in L_2([0,\infty);\mathbb C^n)\cap H^2_{\rm loc}([0,\infty);\mathbb C^n)\,|\,-y''+qy\in L_2([0,\infty);\mathbb C^n)\};\\
& \Dom S_0=\{y\in\Dom S_0^*\,|\, y(0)=y'(0)=0\};\\
& \Dom S=\{y\in\Dom S_0^*\,|\,y(0)=0\};\\
& \mK={\rm Ker\,}S^*_0=\{y\in\Dom S_0^*\,|\,\,-y''+qy=0\},\quad n_\pm^{S_0}={\rm dim\,}\mK= n.
\end{align*}
\smallskip

\noindent$\ast$\,\,\,The M.\,I.\,Vishik decomposition
\begin{equation*}\label{vishik for s}
\Dom S^*_0=\Dom S_0\dotplus S^{-1}\mK\dotplus\mK=\Dom S\dotplus\mK
\end{equation*}
of $y\in\Dom S^*_0$ is
$$
y=y_0+L^{-1}g+h;\quad y_0\in\Dom S_0,\quad g,h\in\mK,
$$
and we have \cite{Vishik,MMM}
\begin{equation}\label{gammas}
    \Gamma_1y=-h,\quad\Gamma_2y=g.
\end{equation}
Since $\dim\mK=n$, there exist exactly $n$ linearly independent
$\mathbb C^n$-valued solutions of the equation $-y''+qy=0$ which
belong to $L_2([0,\infty);\mathbb C^n)$. Take them as columns to
form the matrix $K$. It is a matrix-valued square summable
solution of the same equation. The matrix $K(0)$ is
non-degenerate: if it were degenerate, there would exist a zero
non-trivial linear combination of its columns, and hence an
element $y\in\mK$ with $y(0)=0$. That would mean that $y\in\Dom
S$, which is impossible owing to the Vishik's decomposition, since
$\Dom S\cap\mK=\{0\}$. One can multiply $K$ by $K^{-1}(0)$ and
assume that $K(0)=I$ from the beginning. Let $K_1:=S^{-1}K$ in the
sense that each column of $K_1$ is obtained by applying $S^{-1}$
to the corresponding column of $K$ as a vector-valued solution as
an element of $L_2([0,\infty);\mathbb C^n)$. Since each column of
$K_1$ belongs to $\Dom S$, it should vanish at $x=0$. One has
\begin{equation}\label{gandh}
    g(x)=K(x)c,\quad h(x)=K(x)d
\end{equation}
with some constants $c,d\in\mathbb C^n$. To find them, we use the
fact that $y_0\in\Dom S_0$, so $y_0(0)=0$ and $y_0'(0)=0$. This
can be written as
$$
y_0(0)=y(0)-d=0;\quad y_0'(0)=y'(0)-K_1'(0)c-K'(0)d=0.
$$
Then
$
d=y(0),
$
and the second equality implies
$$
c=(K_1')^{-1}(0)[y'(0)-K'(0)y(0)].
$$
The matrix $K_1'(0)$ is non-degenerate for similar reasons to why
$K(0)$ is: otherwise there would exist a vector $y\in\mK$ such
that $S^{-1}y\in\Dom S_0$, but $\Dom S_0\cap S^{-1}\mK=\{0\}$.
Substituting $c$ and $d$ to the relations \eqref{gandh} and
\eqref{gammas}, we get
\begin{align}\label{Eq Gamma 1}
& \Gamma_1y=-K(x)y(0),\\
\notag
& \Gamma_2y=K(x)(K_1')^{-1}(0)\left[y'(0)-K'(0)y(0)\right].
\end{align}
\smallskip

\noindent{\bf 2.}\,\,\,\ Consider the dynamical system with
boundary control $\alpha^T$ for $S$.

\noindent$\ast$\,\,\,Taking into account (\ref{Eq Gamma 1}), one
can rewrite the system (\ref{Eq 1})--(\ref{Eq 3}) in the form
\begin{align}
\label{Eq 1Q}& u_{tt}-u_{xx}+q(x)u = 0,  && x>0, \,\,\,0<t<T;\\
\label{Eq 2Q}& u|_{t=0}=u_t|_{t=0}=0, && x\gs 0;\\
\label{Eq 3Q}& -Ku\mid_{x=0} = f(t), && 0\ls t\ls T.
\end{align}
Here the corresponding inner and outer spaces are
$\mH=L_2([0,\infty);\mathbb C^n)$ and $\mFT=L_2([0,T];\mK)$, the
solution $u^f(x,t)$ as function of $x$ is supposed to be from
$\Dom S_0^*$ and the differentiation in $t$ is understood in the
sense of differentiating of an $\mH$-valued function.

\def\fsubv{f_{\rm v}}
One can parametrize $f(x,t)=-K(x)f_{\rm v}(t)$ with a
vector-valued function $\fsubv\in\mF_{\rm v}^T:=L_2([0,T];\mathbb
C^n)$. Define the maps $\lambda:\mathbb C^n\to\mK$,
$\lambda:v\mapsto-K(\cdot)v$, and
$$
\Lambda:L_2([0,\infty);\mathbb C^n)\mapsto L_2([0,\infty);\mK),\quad (\Lambda\fsubv)(t)=\lambda(\fsubv(t)),\quad t\in[0,\infty),
$$
as well as its restrictions $\Lambda^T:L_2([0,T];\mathbb C^n)\to
L_2([0,T];\mK)$, $T>0$. Then the system \eqref{Eq 1Q}--\eqref{Eq
3Q} becomes
\begin{align}
\label{Eq 1hat}& u_{tt}-u_{xx}+q(x)u = 0,  && x>0, \,\,\,0<t<T;\\
\label{Eq 2hat}& u|_{t=0}=u_t|_{t=0}=0, && x\gs 0;\\
\label{Eq 3hat}& u\mid_{x=0} = \fsubv(t), && 0\ls t\ls T,
\end{align}
where $\fsubv = (\Lambda^T)^{-1}f$ and the derivative with respect
to the variable $t$ is understood in the same way. An analog of
the control operator for the system \eqref{Eq 1hat}--\eqref{Eq
3hat} can be defined by the equality
$$
(W_{\rm v}^T\fsubv)(\cdot)=u^{\fsubv}(\cdot, T),\quad \fsubv\in \mF_{\rm v}^T.
$$
In \cite{Simonov-preprint} this situation is considered in detail
and it is shown that the solution $u^{\fsubv}$ has the following
representation:
\begin{equation}\label{Eq Repres u^f}
u^{\fsubv}(x,t)=\fsubv(t-x)+\int_x^tw(x,s)\fsubv(t-s)\,ds,\quad x\gs 0,\quad0\ls t\ls T,
\end{equation}
which holds under the agreement that $\fsubv|\,_{t<0}\equiv 0$.
Here $w$ is a continuous matrix-valued kernel which obeys
$w(0,\cdot)\equiv0$. Clearly one has
$$
W^T=W_{\rm v}^T(\Lambda^T)^{-1}.
$$

\noindent$\ast$\,\,\,Let $Y^T_{\rm v}$ denote the reflection
operator in $L_2([0,T];\mathbb C^n)$, $(Y^T\rmv
f\rmv)(x):=f\rmv(T-x)$,  $x\gs 0$. Then the operator
$W_{\rm v}^TY^T_{\rm v}-I$ is a Volterra integral operator in
$L_2([0,T];\mathbb C^n)$, and $W_{\rm v}^TY^T_{\rm v}$ is an
isomorphism. Therefore $W_{\rm v}^T$ is also an isomorphism of
$L_2([0,T];\mathbb C^n)$. The linear set
$$
\dot{\mF_{\rm v}^T}:=(\Lambda^T)^{-1}\dot\mF^T=\{\fsubv\in C^{\infty}([0,T];\mathbb C^n)\,|\,\supp\fsubv\subset(0,T]\}
$$
is dense in $\mF_{\rm v}^T$ and one has $\dot{\mU}^T=W_{\rm v}^T\dot{\mF_{\rm v}^T}$. Consequently, $\overline{\dot{\mU}^T}=W_{\rm v}^T\overline{\dot{\mF_{\rm v}^T}}=W_{\rm v}^TL_2([0,T];\mathbb C^n)=L_2([0,T];\mathbb C^n)$,
$$
\mU^T=L_2([0,T];\mathbb C^n),
$$
and the control operators $W^T_{\rm v}:\mF^T_{\rm v}\to\mU^T$ and $W^T:\mF^T\to\mU^T$ are isomorphisms.

One can show \cite[Theorem 3]{Simonov-preprint} that $W_{\rm v}^T$ is also an isomorphism of $H^2([0,T];\mathbb C^n)$. This means that
\begin{multline*}
\Dom S_0^{*\,T}
=\overline{\dot{\mU}^T}^{S_0^*}
=\overline{\dot{\mU}^T}^{H^2}
=\overline{W_{\rm v}^T\dot{\mF_{\rm v}^T}}^{H^2}
=W_{\rm v}^T\overline{\dot{\mF_{\rm v}^T}}^{H^2}
\\
=W_{\rm v}^T(\{\fsubv\in H^2([0,T];\mathbb C^n)\,|\,\fsubv(0)=\fsubv'(0)=0\})
\\=\{y\in H^2([0,T];\mathbb C^n)\,|\,((W_{\rm v}^T)^{-1}y)(0)=((W_{\rm v}^T)^{-1}y)'(0)=0\}.
\end{multline*}
It is easy to see from \eqref{Eq Repres u^f} that conditions $\fsubv(0)=\fsubv'(0)=0$ are equivalent to $y(T)=y'(T)=0$. We conclude that
\begin{equation}\label{dom s0tstar}
\Dom S_0^{*\,T}=\{y\in H^2([0,T];\mathbb C^n)\,|\,y(T)=y'(T)=0\}.
\end{equation}
One can check now that Condition \ref{C L0 split} is satisfied for $S_0^*$. Indeed, for $a,b\in[0,T]$ such that $0\leqslant a<b\leqslant T$ one has
$$
\Dom S_0^{*\,T}\cap\mU^{ab}=\{y\in H^2([0,T];\mathbb C^n)\,|\,\supp y\subset[a,b],y(T)=y'(T)=0\}.
$$
This linear set is dense in $\mU^{ab}$, and clearly $S_0^{*\,T}(\Dom S_0^{*\,T}\cap\mU^{ab})\subset\mU^{ab}$. Therefore $\mU^{ab}$ is an invariant subspace of $S_0^{*\,T}$. Moreover, if $0<a<b\leqslant T$, then
$$
\Dom S_0^{*\,T}\cap\mU^{ab}=\mathring H^2([a,b];\mathbb C^n),
$$
and for $y\in\Dom S_0^{*\,T}\cap\mU^{ab}$ integrating by parts gives $(S_0^{*\,T}y,y)_{\mH}=\int_a^b(\|y'\|^2+(qy,y))\in\mathbb R$. Hence the part ${S_0^{*\,T}}_{\mU^{ab}}$ is symmetric, which means that Condition \ref{C L0 split} is satisfied.

\medskip
\noindent$\ast$\,\,\,Condition \ref{C5} can also be checked now. We see that $\Dom S_0^{*\,\infty}$ contains $\Dom S_0^{*\,T}$ for all $T>0$, and hence contains $C^{\infty}_{\rm c}([0,\infty);\mathbb C^n)$. The closure of the restriction $S_0^*\upharpoonright C^{\infty}_{\rm c}([0,\infty);\mathbb C^n)$, on the one hand, is contained in $S_0^{*\,\infty}$ and, on the other, coincides with the maximal operator which is $S_0^*$ (this follows from the Povzner--Wienholtz theorem). Therefore $S_0^{*\,\infty}=S_0^*$.
\smallskip

\noindent{$\bf 3.$}\,\,\,Consider the diagonal construction.

\medskip
\noindent$\ast$\,\,\,Choose a partition $\Xi$ of $[0,T]$ of a sufficiently small range $\delta$ and recall that $X^T_s$ cuts off controls to the segment $[T-s,T]$. Take any $f_{\rm v}\in\mF^T_{\rm v}$ with $f=\Lambda^Tf_{\rm v}\in\mF^T$ and compose the sums (\ref{Eq sums}) for the operators $W^T$ and $W^T_{\rm v}$:
\begin{multline*}\label{Eq diag W}
D_{W^T}^\Xi f
=\sum\limits_{k=0}^N\Delta P^T_kW^T\Delta X^T_kf
=\sum\limits_{k=0}^N\Delta P^T_kW^T_{\rm v}\underbrace{(\Lambda^T)^{-1}\Delta X^T_k\Lambda^T}_{\Delta X^T_{{\rm v}_k}}f_{\rm v}
\\
=\sum\limits_{k=0}^N\Delta P^T_kW^T_{\rm v}\Delta X^T_{{\rm v}_k}f_{\rm v}
=D_{W^T_{\rm v}}^\Xi f_{\rm v},
\end{multline*}
where $\{X^T_{{\rm v}_s}\}_{0\ls s\ls T}$ is the nest of projections in $\mF^T_{\rm v}$ on $\mF^T_{{\rm v}_s}=(\Lambda^T)^{-1}\mF^T_s$ and $\Delta X^T_{{\rm v}_k}:=X^T_{{\rm v}_{s_k}}-X^T_{{\rm v}_{s_{k-1}}}=(\Lambda^T)^{-1}\Delta X^T_k\Lambda^T$. Thus the sums always converge simultaneously and we can show the existence of the diagonal of $W^T_{\rm v}$ with respect to the nest $\{\mF^T_{{\rm v}_s}\}_{0\leqslant s\leqslant T}$. It also follows that if the diagonals exist, they are related by
\begin{equation*}
D_{W^T}=D_{W^T_{\rm v}}(\Lambda^T)^{-1}.
\end{equation*}

Consider the $k$-th summand. Taking into account the fact that $\Delta P^T_k$ cuts off functions to the segment $[s_{k-1},s_k]$ and $\Delta X^T_{{\rm v}_k}$ to the segment $[T-s_k,T-s_{k-1}]$, by the representation (\ref{Eq Repres u^f}) one has
\begin{multline*}
(\Delta P^T_kW^T_{\rm v}\Delta X^T_{{\rm v}_k}f_{\rm v})(x)
\\=
\left\{
\begin{array}{cl}
f_{\rm v}(T-x)+\int_x^{s_k}w(x,s)f_{\rm v}(T-s)ds,&x\in
[s_{k-1},s_k],
\\
0,&x\in[0,T]\backslash[s_{k-1},s_k].
\end{array}
\right.
\end{multline*}
Denote $w_k(x):=\int_x^{s_k}w(x,s)f_{\rm v}(T-s)ds$, $x\in[s_{k-1},s_k]$, $k=1,...,n$, and $\omega:=\max_{\{(x,t)\vert t\in[0,T],x\in[0,t]\}}\|w(x,t)\|_{\matr}^2$. Estimates give:
\begin{multline*}
\|w_k\|^2_{L_2([s_{k-1},s_k];\mathbb C^n)}
\leqslant\omega^2
\int_{s_{k-1}}^{s_k}\left(\int_{x}^{s_k}\|f_{\rm v}(T-s)\|_{\mathbb C^n}ds\right)^2dx
\\
\leqslant\delta\omega
\left(\int_{T-s_k}^{T-s_{k-1}}\|f_{\rm v}\|_{\mathbb C^n}\right)^2
\leqslant\delta^2\omega\int_{T-s_k}^{T-s_{k-1}}\|f_{\rm v}\|_{\mathbb C^n}^2.
\end{multline*}
Then
\begin{equation*}
\|D_{W^T_{\rm v}}^\Xi f_{\rm v}(\cdot)- f_{\rm v}(T-\cdot)\|^2_{\mathscr
H}
=\sum\limits_{k=0}^N\|\omega_k\|^2_{L_2([s_{k-1},s_k];\mathbb C^n)}\ls
\delta^2\omega\|f_{\rm v}\|^2_{\mathscr F^T_{\rm v}}.
\end{equation*}
As a result we conclude that the sums converge as $\delta\to 0+$ in norm, i.\,e., the diagonal $D_{W^T_{\rm v}}=\int_{[0,T]}dP_sW^T_{\rm v}\,dX^T_{{\rm v}_s}ds$ converges in the strong sense and acts from $\mathscr F^T_{\rm v}$ to $\mathscr H$ by the rule
\begin{equation*}
(D_{W^T_{\rm v}}f_{\rm v})(x)=f_{\rm v}(T-x),\quad x\in[0,T].
\end{equation*}
As we mentioned above, the diagonal $D_{W^T}=D_{W^T_{\rm v}}(\Lambda^T)^{-1}$ also exists and is an isomorphism of $\mF^T$ and $\mU^T$, because clearly both $\Lambda^T$ and $D_{W^T_{\rm v}}$ are isomorphisms. Thus Condition \ref{C2} is satisfied. A remarkable fact is that the diagonal $D_{W^T\rmv}$ does not depend on $q$.
\smallskip

\noindent{$\bf 4.$}\,\,\,To check Condition \ref{C3} we need to find $\Dom\tilde S_0^{*\,T}=\Phi^T\Dom S_0^{*\,T}$, where $\Phi^T=\tilde Y^T\Phi_{D_{W^T}^*}$. Since the operators $(\Lambda^T)^{-1}$ and $D_{W^T_{\rm v}}$ commute, one has
\begin{equation*}
|D^*_{W^T}|^2=D_{W^T}D^*_{W^T}=D_{W^T_{\rm v}}(\Lambda^T)^{-1}((\Lambda^T)^{-1})^*D^*_{W^T_{\rm v}}=((\Lambda^T)^*\Lambda^T)^{-1}=[(\lambda^*\lambda)^{-1}],
\end{equation*}
where $[\cdot]$ denotes the operator of multiplication by the constant matrix $(\lambda^*\lambda)^{-1}$. For $f=\lambda f_{\rm v}$, $g=\lambda g_{\rm v}$ one has
\begin{multline*}
(\lambda^*\lambda f_{\rm v},g_{\rm v})_{\mathbb C^n}=(\lambda f_{\rm v},\lambda g_{\rm v})_{\mK}=(Kf_{\rm v},Kg_{\rm v})_{\mH}
\\
=
\sum_{i,j=1}^nf^i_{\rm v}g^j_{\rm v}(k_i,k_j)_{\mH}=(G_Kf_{\rm v},g_{\rm v})_{\mathbb C^n},
\end{multline*}
where $G_K$ is the Gram matrix of the system of vectors $k_1(x)$, ..., $k_n(x)$, which are the columns of the matrix $K(x)$, $(G_K)_{ij}=(k_i,k_j)_{\mH}$. Therefore $\lambda^*\lambda=G_K$ and $|D^*_{W^T}|=[G_K^{-\frac12}]$. Then, since $(\lambda^{-1})^*=\lambda G_K^{-1}$,
\begin{equation*}
\Phi_{D^*_{W^T}}=D^*_{W^T}(|D^*_{W^T}|)^{-1}=[(\lambda^{-1})^*]D^*_{W^T_{\rm v}}[G_K^{\frac12}]=[\lambda G_K^{-1}]D^*{W^T_{\rm v}}
\end{equation*}
with the same meaning of $[\cdot]$ as a `pointwise' constant operator, and
\begin{equation*}
\Phi^T=\tilde Y^T\Phi_{D^*_{W^T}}=\tilde Y^T[\lambda G_K^{-1}]D^*_{W^T_{\rm v}}=[\lambda G_K^{-\frac12}]
\end{equation*}
(which is indeed a unitary operator from $\mF^T_{\rm v}$ to $\mF^T$), because both $\tilde Y^T$ and $D^*_{W^T_{\rm v}}$ act as reflection operators. Since $\lambda G_K^{-\frac12}$ is an isomorphism of $\mathbb C^n$ and $\mK$, it follows from \eqref{dom s0tstar} that
\begin{equation*}
\Phi^T\Dom S_0^{*\,T}=\{\tilde u\in H^2([0,T];\mK)\,\vert\,\tilde u(T)=\tilde u'(T)=0\},
\end{equation*}
which shows that Condition \ref{C3} is satisfied.

\noindent{$\bf 5.$}\,\,\,Consider
\begin{multline*}
Q^T
=\Phi^TS_0^{*\,T}(\Phi^T)^*+\dtau
\\
=
[\lambda G_K^{-\frac12}]
\left(
-\dtau+[q(\tau)]
\right)
[G_K^{\frac12}\lambda^{-1}]+\dtau
=[\lambda G_K^{-\frac12}q(\tau)G_K^{\frac12}\lambda^{-1}].
\end{multline*}
The operator $\lambda G_K^{-\frac12}q(\tau)G_K^{\frac12}\lambda^{-1}$ is bounded in $\mK$ (for a.\,e. $\tau$) and
\begin{equation*}
\|Q^T(\tau)\|_{\fB(\mK)}\ls\|\lambda\|_{\fB(\mathbb C^n,\mK)}\|\lambda^{-1}\|_{\fB(\mK,\mathbb C^n)}\|G_K^{-\frac12}\|_{\matr}\|G_K^{\frac12}\|_{\matr}\|q(\tau)\|_{\matr}.
\end{equation*}
It follows that $Q^T\in L_{\infty}([0,T];\fB(\mK))$, which means that Condition \ref{C4} is satisfied.

We have shown that Conditions \ref{C L0 split}--\ref{C5} hold for $S_0$ and for any symmetric operator unitarily equivalent to $S_0$, hence the proof is complete.
\end{proof}

\subsubsection*{Comments}

\bul In applications, constructing a Schr\"odinger model of an operator provides a way for solving inverse problems.
For a wide class of problems the connecting operator $C^T$ is determined by the inverse data \cite{B Obzor IP 97,B DSBC IP 2001,B UMN}.
Owing to this, given appropriate inverse data, it is possible to realize triangular factorization (\ref{Eq sqrt C^T}) and perform a procedure that produces the model $\tilde L_0$ and thus determines the `potential' $Q$.
In view of the invariant character of the wave model construction, the appropriate data can be anything which determines the operator $L_0$ up to unitary equivalence.
For instance, the characteristic function of $L_0$ is a valid data.

\bul The local boundedness of $Q$ is typical for one-dimensional inverse problems, whereas the case of unbounded $Q$ corresponds to multidimensional settings.
To generalize the above scheme to this case would be an interesting and important task.
However, the necessity in Theorem \ref{T3} may be a dificult matter.

\bul Not much is said in the paper about the eikonal operator $E$, which in essence is a background for the wave model.
To construct the latter, we determine $E$ via the systems $\alpha^T$, $T>0$, and diagonalize it by the Fourier transform associated with diagonals of operators $W^T$, which control the wave propagation in $\alpha^T$.

\bigskip
\noindent
{\bf Keywords:}
Triangular factorization, nest theory, functional model, inverse problem, characterization, matrix Schr\"odinger operator.

\smallskip
\noindent
{\bf MSC:}
47A45, %Canonical models for contractions and nonselfadjoint linear operators
47A46, %Chains (nests) of projections or of invariant subspaces, integrals along chains, etc.
47A68, %Factorization theory (including Wiener-Hopf and spectral factorizations) of linear operators
47B25, %Linear symmetric and selfadjoint operators (unbounded)
47B93, %Operators arising in mathematical physics
35R30. %Inverse problems for PDEs

\end{document}